\documentclass[
  aps,
  prb,
  reprint,
  onecolumn,
  amsfonts,
  amssymb,
  amsmath,
]{revtex4-1}

\usepackage{bm}
\usepackage{graphics}
\usepackage{graphicx}
\usepackage{color}
\usepackage[hidelinks]{hyperref}
\usepackage[svgnames]{xcolor}

\def\red#1{\textcolor{black}{#1}}
\def\blue#1{\textcolor{black}{#1}}

\newcommand{\bolds}{\mathbf {s}}
\newcommand{\boldv}{\mathbf {v}}

\begin{document}

\title{Friction-enhanced lifetime of bundled quantum vortices}

\author{Luca Galantucci}
\affiliation{Istituto per le Applicazioni del Calcolo `M. Picone’,
IAC-CNR, Via dei Taurini 19, 00185 Roma, Italy}
\affiliation{Joint Quantum Centre (JQC) Durham--Newcastle, and
School of Mathematics and Statistics, Newcastle University,
Newcastle upon Tyne, NE1 7RU, United Kingdom}

\author{Giorgio Krstulovic}
\affiliation{Université Côte d'Azur, Observatoire de la Côte d'Azur, CNRS,
Laboratoire Lagrange, Boulevard de l'Observatoire CS 34229 - F 06304 NICE Cedex 4, France}

\author{Carlo~F. Barenghi}
\affiliation{Joint Quantum Centre (JQC) Durham--Newcastle, and
School of Mathematics and Statistics, Newcastle University,
Newcastle upon Tyne, NE1 7RU, United Kingdom.}

\date{\today}


\begin{abstract}
We show that a toroidal bundle of quantized vortex rings
in superfluid helium generates a large-scale wake in the normal 
fluid which reduces the 
overall friction experienced by the bundle, thus greatly 
enhancing its lifetime, as observed in experiments.
This collective effect is similar to the drag reduction observed in
systems of active, hydrodynamically cooperative agents 
such as bacteria in aqueous suspensions,
fungal spores in the atmosphere and cyclists in pelotons. 
\end{abstract}


\maketitle

\section{Introduction}
Some physical systems consist of components
which interact with each others not only directly but also indirectly
by changing the common background, leading to remarkable collective effects
such as drag reduction.
Examples are aqueous suspensions of self-propelled bacteria 
\cite{Lopez2015,Martinez2020,Guo2018bacteria}, fungal spores
\cite{Roper2010}, road racing cyclists in the peloton 
\cite{Blocken2018,Belden2019}, and particles 
trapped inside an optical vortex \cite{Reichert2004,Grujic2007}.
Here we report a similar collective effect for
quantized vortex rings, 
fundamental nonlinear excitations of superfluid helium.
Vortex rings are generated in the laboratory by injecting electrons 
\cite{Gamota1973,Walmsley2008,Walmsley2014}, forcing liquid helium through
orifices \cite{Guenin1978} or moving a grid \cite{Bradley2005}. 
At sufficiently low temperatures, a single, isolated superfluid
vortex ring of radius $R$
is an Hamiltonian object \cite{Barenghi2009} traveling
at constant energy ($\propto R$) and velocity ($\propto 1/R$).
At higher temperatures, liquid helium has a two-fluid nature: 
thermal excitations (phonons and rotons) form a viscous gas called
the normal fluid which interacts with quantized vortices via a mutual 
friction force. Because of this friction, a superfluid vortex ring 
moving in a quiescent normal fluid loses energy, shrinks, speeds up 
and vanishes.

Here we show that the dynamics of a sufficiently compact toroidal bundle 
of {\it many} vortex rings is remarkably different: 
besides interacting directly with each others in a peculiar leapfrogging 
fashion, the vortex rings also interplay indirectly by
modifying the common normal fluid background. This reduces
drastically the total friction so that
the bundle remains coherent as if the normal fluid were almost absent, 
resulting in an enhanced lifetime. Such unusual long life of superfluid 
vortex bundles has been observed in experiments
\cite{Borner1981,Borner1983,Borner1985}, but never explained until now.

In experiments \cite{Borner1981,Borner1983,Borner1985}, large-scale
vortex ring structures identified as vortex bundles were generated 
by forcing liquid helium out of a cylindrical tube.
Position and translational velocity of the structures
were measured acoustically, together with the spatial distributions 
of superfluid and normal fluid circulations.
It was found that, over a wide temperature range ($1.3<T<2.15~\rm K$,
corresponding to the superfluid fraction changing from
$96 \%$ to $0.13 \%$)
the bundles remained relatively compact, conserving their initial shape and 
moving at constant speed over distances of the order of 7 times their 
initial diameter.
The measured superfluid and normal fluid circulations were both
of the order of $10^3 \kappa$, where $\kappa$ is the quantum of circulation of one single ring. This fact suggests that such fluid structures consisted of a bundle of $\approx 10^3$ superfluid vortex rings 
embedded in a normal fluid vortex structure of the same circulation,
traveling together across the apparatus.

\begin{figure*}[t!]
       \centering
       \includegraphics[width=.95\linewidth]{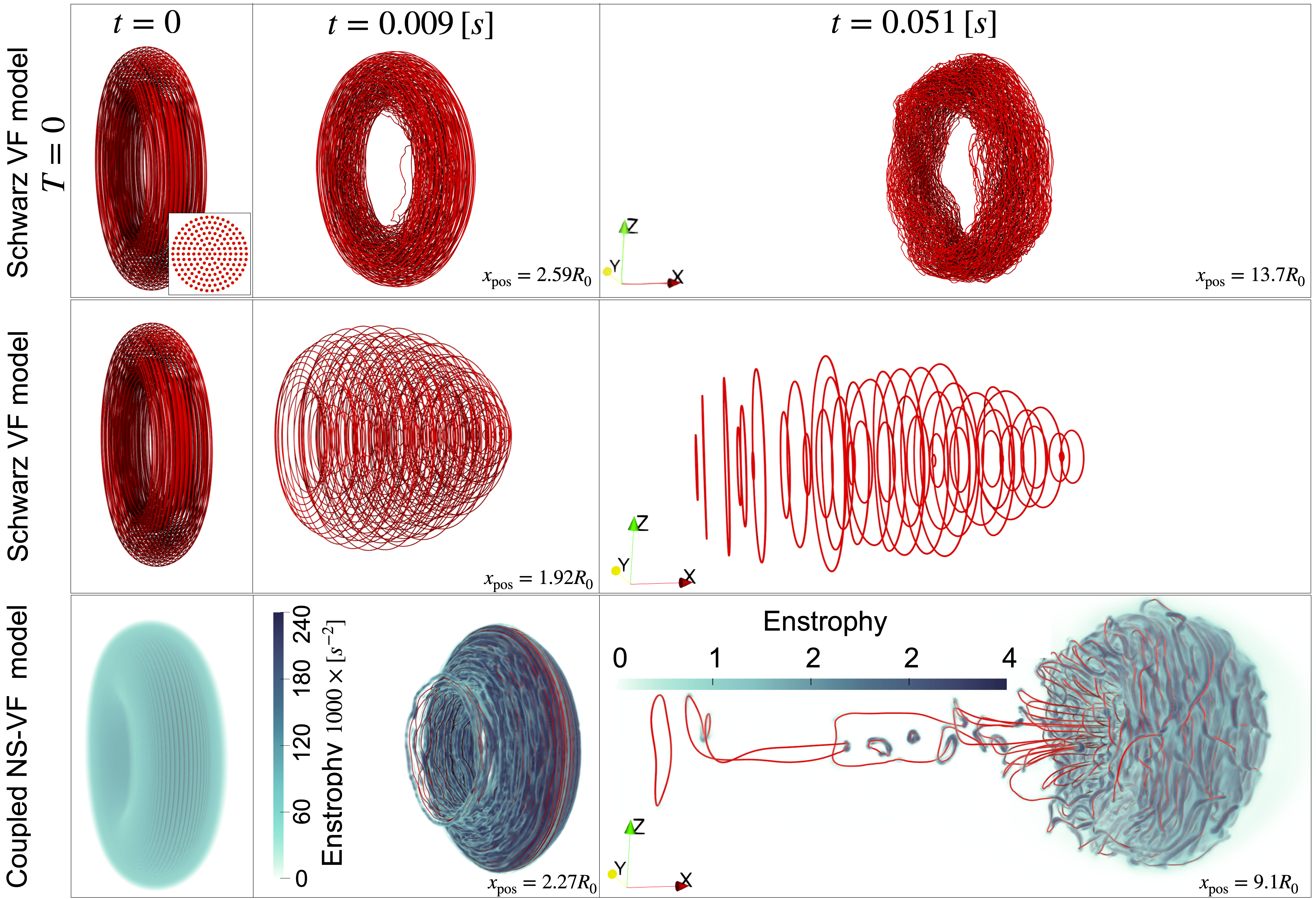}
       \caption{
       Evolution of vortex bundle with an initial central radius 
       $R_0=1.2 \times 10^{-2}~\rm cm$ traveling along the x-direction
       at $t=0$, (left column), $t=0.009~\rm s$ (middle column) 
       and $t=0.051~\rm s$ (right column).
       Vortex lines are displayed in red and the normal fluid's 
       enstrophy in blueish colours. The inset in the top left panel 
       displays the initial cross section of the bundle. 
       Top and middle rows show numerical simulations  
       of Schwarz's VF model at $T=0$ and $T=1.95~\rm K$ respectively. 
       The distance travelled by the bundle (where it remains 
       coherent) is denoted by $x_{\rm pos}$.
	Results obtained using the coupled NS-VF model are shown in the 
        bottom row. 
        }
       \label{fig:BundleVis3D}
\end{figure*}

\section{Model and numerical experiment}
Our model builds on the Vortex Filament (VF) theory of Schwarz 
\cite{Schwarz1988}, a widely used approach
\cite{Araki2002,baggaley-laurie-barenghi-2012}
which describes vortex lines as space curves $\bolds(\xi,t)$ of 
infinitesimal thickness moving according to
\begin{equation}
\dot{\bolds}(\xi,t)=\frac{\partial \bolds}{\partial t}=
\boldv_{\rm s}+\alpha \bolds' \times \boldv_{\rm ns}
-\alpha' \bolds' \times (\bolds' \times \boldv_{\rm ns}),
\label{eq:Schwarz}
\end{equation}
where $\bolds'=\partial\bolds/\partial \xi$,
$\boldv_{\rm ns}=\boldv_{\rm n}-\boldv_{\rm s}$ at $\bolds$, 
$\alpha$ and $\alpha'$ are temperature-dependent friction
coefficients \cite{Donnelly1998}, 
$\boldv_{\rm n}$ is the normal fluid velocity at 
$\bolds$, and $\boldv_s$ is the superfluid velocity induced at $\bolds$ by
the entire vortex configuration $\cal L$ via 
\begin{equation}
\boldv_{\rm s}({\bf s},t)=\frac{\kappa}{4 \pi} \oint_{\cal L}
\frac{\bolds_1'(\xi_1,t) \times ({\bf s}-\bolds_1(\xi_1,t))}
{\vert {\bf s} -\bolds_1(\xi_1,t)\vert^3 }d\xi_1 .
\label{eq:BS}
\end{equation}
The original VF model consists of Eqs.~(\ref{eq:Schwarz}) - 
(\ref{eq:BS}) and an algorithm to perform vortex reconnections.
Its limitation is that the normal fluid velocity $\boldv_n$
is imposed {\it a priori}, neglecting the back-reaction of the superfluid 
vortex lines on $\boldv_n$.  Recent experiments 
\cite{guo-etal-2010,mastracci-etal-2019} suggest that normal fluid
wakes may form behind each individual vortex line.
To account for this effect, which is crucial to understand quantized
vortex bundles, we couple Eqs.~(\ref{eq:Schwarz}) and (\ref{eq:BS}) 
self-consistently with the Navier-Stokes equations for $\boldv_n$ supplemented with a mutual friction force $\mathbf{F}_{\rm ns}$:
\begin{eqnarray}
\displaystyle
\frac{\partial \boldv_{\rm n}}{\partial t} + \left ( \boldv_{\rm n}\cdot \nabla \right )\boldv_{\rm n}   = 
-\frac{1}{\rho} \nabla p_{\rm n} &+& \nu_{\rm n} \nabla^2 \boldv_{\rm n} +\frac{\mathbf{F}_{\rm ns}}{\rho_{\rm n}}
\label{eq:NS} \\
\mathbf{F}_{\rm ns}=\oint_\mathcal{L}\mathbf{f}_{\rm ns}(\mathbf{s})\delta({\bf x}-\mathbf{s})\mathrm{d}\xi\,\,, &  &  \nabla \cdot \boldv_{\rm n}=0 \label{eq:incompr} \; .
\end{eqnarray}
Here $\mathbf{f}_{\rm ns}$ is the local friction per unit length,
$\rho= \rho_{\rm n} + \rho_{\rm s}$ is the total density of liquid helium, 
$\rho_{\rm n}$ and $\rho_{\rm s}$ are respectively the normal fluid
and superfluid densities, $p_{\rm n}$ is the effective pressure, 
and $\nu_{\rm n}$ the kinematic viscosity of the normal fluid.
We refer to Eqs.~(\ref{eq:Schwarz}) - (\ref{eq:incompr}) as
the coupled Navier-Stokes Vortex Filament (NS-VF) model
\cite{Galantucci2020}. Further details and
comparisons with previous approaches 
\cite{kivotides-barenghi-samuels-2000,galantucci-sciacca-barenghi-2015,kivotides-2018,yui-etal-2019} are in 
described in Appendices~\ref{App:method} and \ref{App:Params}.
 
The initial condition of our numerical experiments consists
of a concentric bundle of $N=169$ circular vortex rings placed inside 
a torus of outer radius $R_0=1.2\times 10^{-2} \text{cm}$ and inner radius 
$a=3.4\times 10^{-3}\text{cm}$.  These initial rings are distributed 
in a regular hexagonal lattice over the torus cross-section 
(top left panel of Fig.~\ref{fig:BundleVis3D}), 
corresponding to solid-body rotation within the torus. 
The exact initial vortex configuration does not play a fundamental role,
as we obtain the same numerical results with vortices arranged randomly within the toroidal 
geometry.
For comparison, we also study a smaller bundle 
($R_0=2.3\times 10^{-2}\text{cm}$, $a=2.9\times 10^{-3}\text{cm}$, $R/a=8$) 
with only $N=37$ vortex rings. Although for practical computational
reasons we have about 10 times less rings than in experiments, the
radius ratio $R_0/a = 3.5$ of the larger bundle is essentially the same as in \cite{Borner1985}.

\section{Results}
\begin{figure}[h]
       \centering
       \includegraphics[width=1.00\linewidth]{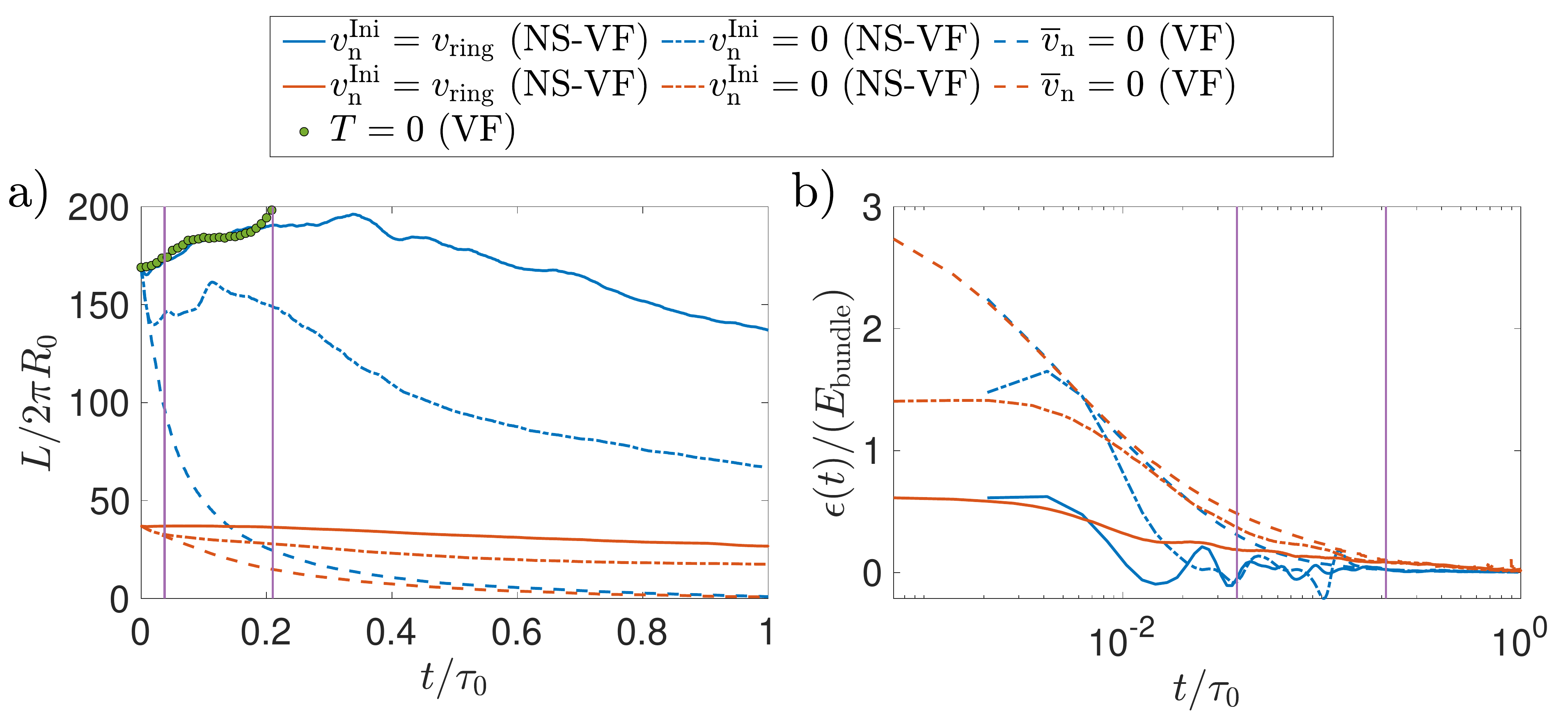}
	\caption{ Temporal evolution of vortex bundle's 
	total length $L(t)$ \textbf{(a)} and energy dissipation $\epsilon(t)$ \textbf{(b)} 
        at $T=1.95~\rm $K of bundles with $N=169$ (blue) and $N=37$ (red) rings.
	Plots compare Schwarz's VF model (dashed lines) and 
        coupled NS-VF model with (solid lines) and without (dot-dashed lines) 
        an initial normal fluid vortex ring. 
	$T=0$ temporal evolution is shown by green dots. The two vertical lines correspond to times 
        $t=0.009~\rm s$ and $t=0.051~\rm s$ represented 
        in Fig.~\ref{fig:BundleVis3D}.  
        The time scale $\tau_0$ is the lifetime of a single quantum vortex 
        ring of initial radius equal to $R_0$ at $T=1.95~\rm K$ 
        ($\tau_0=0.2444~\rm s$ and $\tau_0=0.9041~\rm s$ for the $N=169$
        and the $N=37$ bundle, respectively). 
        }
\label{fig:BundleLengthAndEnergy}
\end{figure}

Firstly we perform simulations at temperature $T=0$:
the normal fluid and the friction are
absent (vortices hence move purely according 
to the Biot-Savart law, Eq.~(\ref{eq:BS})). 
We find that the vortex bundle preserves its shape
for a long time as displayed in Fig.~\ref{fig:BundleVis3D} (top). 
The radii $R$ and $a$ remain almost constant during the computed
evolution, while the bundle travels a distance 
$D \, {\approx} \, 14 \, R_0$ in the $x$ direction; this is in
quantitative agreement with experiments \cite{Borner1985}
in the low temperature range
($T=1.3~\rm K$, corresponding to 96 \% superfluid fraction).
In the initial stage, each vortex rings
leapfrogs around and inside the others, until reconnections occur,
triggering Kelvin waves, as illustrated in Fig.~\ref{fig:BundleVis3D} 
\cite{Wacks2014}. Presence of Kelvin waves implies
a small increase of the total vortex length 
$L=\oint_{\cal L}\! \mathrm{d}\xi$ 
(green dots in Fig.~\ref{fig:BundleLengthAndEnergy}(a)).
Only at much later times (not shown), the bundle slowly starts 
losing coherence.

Secondly, we study the bundle's evolution at
$T=1.95~\rm K$ (corresponding to $\rho_{\rm n} \approx \rho_{\rm s}$). 
Using Schwarz's VF model, we observe that,  
travelling in the normal fluid imposed at rest, the bundle
spreads spatially in the direction of motion, rapidly 
losing its coherence by leaving vortices behind 
(see Fig.~\ref{fig:BundleVis3D} (middle)). 
The rapid decay of the total vortex length is clear in 
Fig.~\ref{fig:BundleLengthAndEnergy} (a, dashed blue line):
by $t=0.009~\rm s$ and $t=0.051~\rm s$, $L$ has decreased to 
$60\%$ and $10\%$ of its original value respectively,
in stark disagreement with experiments. Essentially, the bundle
disassembles into isolated vortex rings which shrink
in a time interval comparable to the lifetime 
$\tau_0=0.2444~\rm s$ of a single vortex ring of initial radius equal 
to $R_0$ at the same temperature $T=1.95~\rm K$.

We observe a totally different behavior if we use the more realistic 
coupled NS-VF model \cite{Galantucci2020} accounting for the evolving 
$\boldv_{\rm n}$.  As initial condition for $\boldv_{\rm n}$ we choose 
a large-scale toroidal vortex-ring of outer radius $R_0$, inner radius $a$ 
and circulation $N \kappa$ (i.e. matching the superfluid bundle circulation), 
with a Gaussian distribution of vorticity within the toroidal core (see 
Fig.~\ref{fig:BundleVis3D} (bottom, left)). 
This is probably a fair approximation to
the physical reality of the experiment: as liquid helium is pushed out
of the nozzle, a normal fluid vortex ring with the same circulation of the 
superfluid vortex bundle is indeed observed \cite{Borner1985}. 
We find that the vortex bundle does not decay, 
but remains coherent and travels a significant distance $D$
compared to its diameter ($D \approx 15 R_0$) ,
in agreement with experiments \cite{Borner1985}.
The coherence of the coupled normal fluid -- superfluid vortex structure
can be appreciated in Fig.~\ref{fig:BundleVis3D} (bottom), where
the normal fluid enstrophy density, $|\bm{\omega}_{\rm n}|^2$, is shown 
(bluish colors) alongside the superfluid quantized vortices (red lines).
We observe that the 
radial distribution of the vorticity is broader compared to the initial condition, 
also filling the central 
region of the torus: this is
consistent with the experimental report that 
large-scale helium vortex rings have a less sharp vorticity 
distribution than vortex rings in classical 
fluids \cite{Borner1983,Borner1985}.

Remarkably, while under Schwarz's VF evolution
the total vortex length $L$ rapidly decays, 
under coupled NS-VF evolution $L$ remains almost constant,
see Fig.~\ref{fig:BundleLengthAndEnergy}(a),
similarly to what happens for $T=0$.
This effect is not simply the consequence of the initially
imposed normal fluid ring. We have indeed performed 
NS-VF simulations with an initially quiescent normal fluid 
(dot-dashed lines in Fig.~\ref{fig:BundleLengthAndEnergy}).
We have found that during a 
short initial stage ($t<0.02 \tau_0$), 
the coupled NS-VF model follows the rapid decay of the 
Schwarz's VF model, but after this short transient, the superfluid 
vortex bundle creates normal fluid vortex structures which prevent
the rapid decay of the bundle. 
The evolution of the smaller vortex bundle ($N=37$ rings)
is similar, as shown by the
red curves in Fig.~\ref{fig:BundleLengthAndEnergy}. 
The larger and faster spatial spreading of the initial compact structure as 
the energy saving mechanism is less efficient, is reminiscent of the behaviour observed in 
active matter systems \cite{Trenchard2016}.

\subsection{Dissipation reduction via hydrodynamic cooperation}
The normal fluid vortex structures generated by the back--reaction 
of the superfluid vortex rings are 
similar to
vorticity injection
in ordinary viscous fluids by point-like active agents ({\it e.g.} 
solid particles in classical turbulence \cite{gualtieri_picano_sardina_casciola_2015}), 
suggesting that a superfluid can be seen as a peculiar type of 
active fluid.
\begin{figure}[ht]
       \centering
       \includegraphics[width=.45\linewidth]{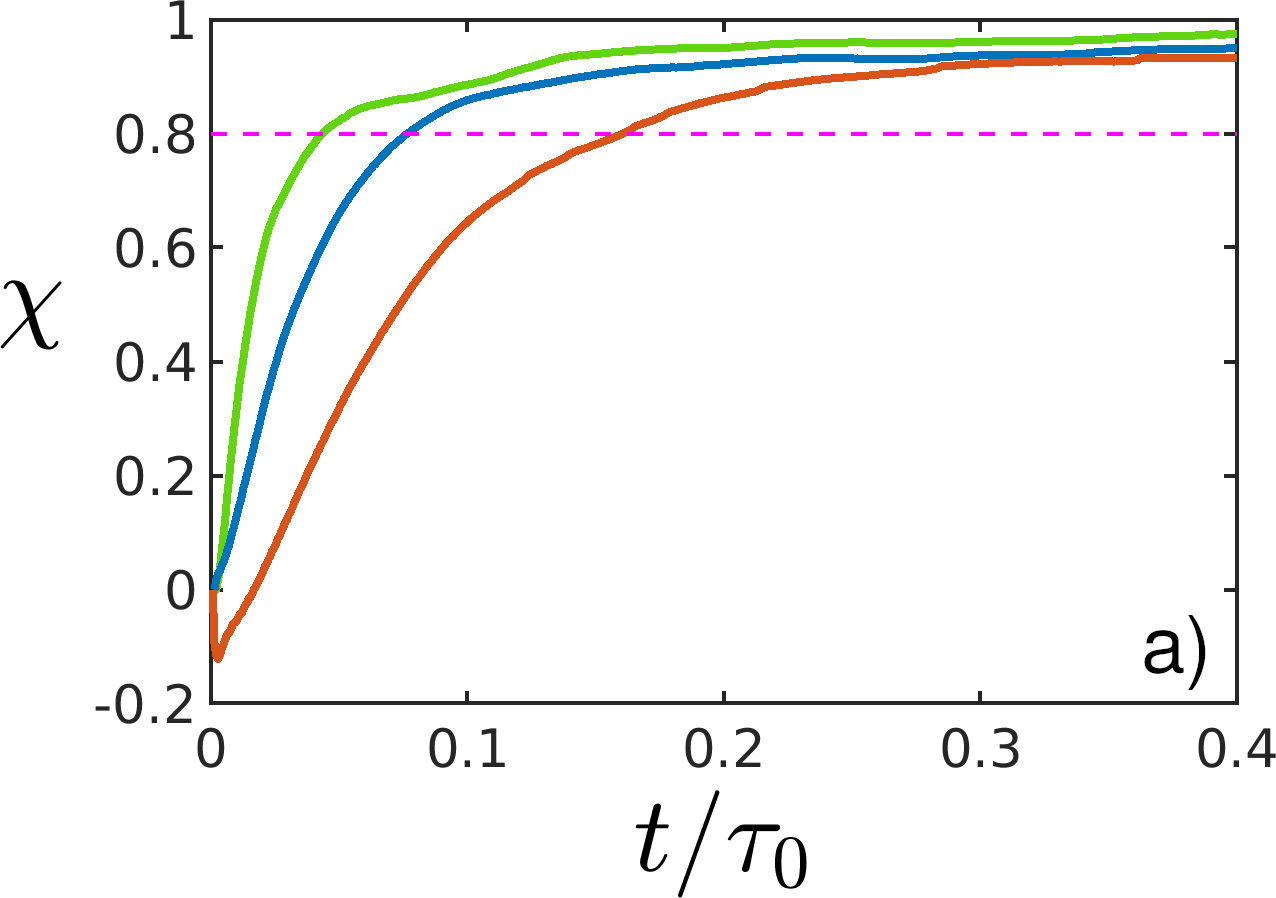}
       \hspace{0.025\textwidth}
       \includegraphics[width=.45\linewidth]{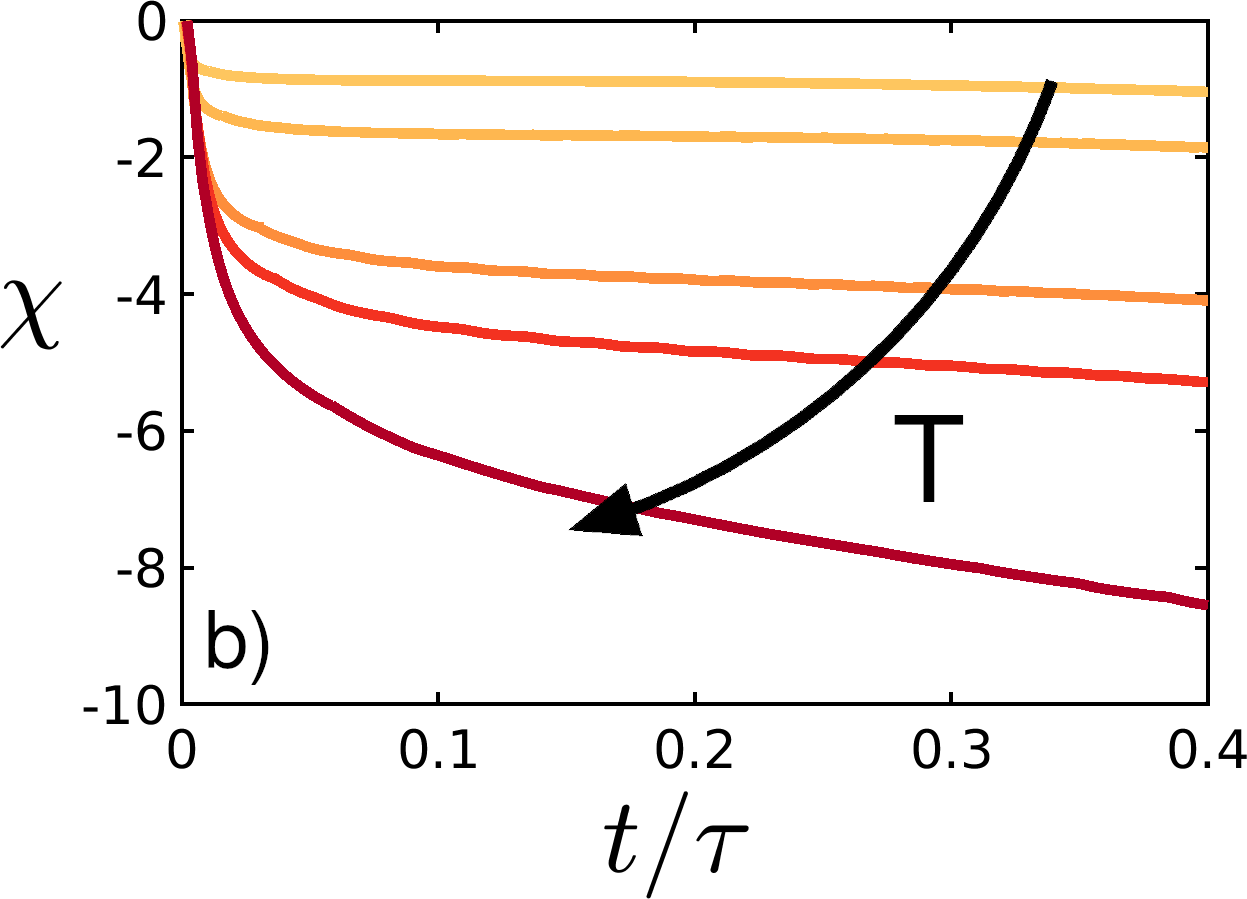}      
       \caption{
       \textbf{a)}: Temporal evolution of dissipation reduction $\chi$ for a superfluid vortex bundle 
	with a number of vortices $N=37$, outer radius $R_0=2.3\times 10^{-2}\text{cm}$ and inner radius 
	$a=2.9\times 10^{-3}\text{cm}$ (green curve, $\ell/\bar{\sigma}=1.7$), $a=5.8 \times 10^{-3} \text{cm}$ 
	(blue curve, $\ell/\bar{\sigma}=3.4$) and 
	$a=1.2 \times 10^{-2} \text{cm}$ (red curve, $\ell/\bar{\sigma}=6.8$). 
	Temperature is $T=1.95~\rm $K. The horizontal dashed magenta line indicates $80\%$ of dissipation reduction.
        Time scale $\tau_0$ as in Fig.~\ref{fig:BundleLengthAndEnergy}.
       \textbf{b)}: Temporal evolution of $\chi$ for an isolated vortex ring of initial radius 
       $R_0=7.6 \times 10^{-3}\text{cm}$, moving in an initially quiescent normal fluid at
       temperatures $T=1.7,\,1.8,\,1.95,\,2.0,\,2.1~\rm K$ (from yellow to red);
       time $t$ is normalized by the vortex ring lifetime $\tau$.}
       \label{fig:dissip.red}
\end{figure}
The mutual friction force per unit length $\mathbf{f}_{ns}$
is a function of the local relative velocity 
$\dot{\bolds}-\boldv_{\rm n}$ 
between the vortex line and the local normal fluid velocity.
If the coupling between superfluid and normal fluid is sufficiently
strong \blue{and the intervortex distance is sufficiently small 
(so that vortices can benefit from the normal fluid stirring performed by other vortices)},
$|\dot{\bolds}-\boldv_{\rm n}|\rightarrow 0$, 
reducing the drag and slowing down (even halting) the decay of 
the combined normal 
fluid - superfluid vortex structure.

To \blue{characterise} the dissipation reduction \blue{arising from the interaction
between vortices and normal fluid}, 
we compute the dissipation of superfluid kinetic energy 
%
\begin{equation}
\epsilon(t) = \oint_{\cal L} \! \mathbf{f}_{\rm ns}(\bolds)
\cdot \dot{\bolds}(\xi,t)\, \mathrm{d}\xi,
\end{equation}
normalized by $\rho_{\rm n}\kappa^2N^2L(t)$, and report it in 
Fig.~\ref{fig:BundleLengthAndEnergy}(b). 
Schwarz's VF model (dashed lines) is compared to
the coupled NS-VF model with and without an initial normal fluid ring 
(solid and dot-dashed lines respectively). It is clear that in 
the coupled model the dissipation is substantially reduced compared 
to Schwarz's VF model (note that in the VF model, the decrease of 
friction at large times is related to the small number of distant vortices 
remaining in the system leading to $\dot{\bolds} \rightarrow 0$). 
%

\blue{Two concurring mechanisms are likely to be responsible for this observed reduced dissipation in
the coupled model: the coupling itself, which reduces the velocity difference $\dot{\mathbf{s}}-\mathbf{v}_n$
between a {\it single} vortex ring and the normal fluid, and the collective hydrodynamic cooperation, 
where vortices benefit from the normal fluid stirring performed 
by other vortices. }
\blue{To determine the relevance of collective effects, 
we study the impact of the average intervortex distance on the dissipation reduction,
by numerically simulating the dynamics of bundles with different initial
inner radii $a$ and computing the dissipation reduction $\chi$ with respect to the initial condition,
defined as $ \chi (t) = (\epsilon(0)-\epsilon(t))/(\epsilon(0))$. 
The initial condition of the normal fluid is quiescent, as this allows to better appreciate 
the stirring of the normal fluid performed by superfluid vortices and the consequent dissipation reduction.
The temporal evolution of $\chi$ is illustrated in
Fig.~\ref{fig:dissip.red}(a) where it
clearly emerges that the dissipation reduction is more efficient when $a$ is smaller. 
The time interval which the system requires 
to reduce the initial dissipation $\epsilon(0)$ by $80\%$ (magenta dashed line in Fig.~\ref{fig:dissip.red}(a)), 
is almost proportional to $a$.}
\blue{This observed less efficient dissipation reduction as $a$ increases 
is determined by the following factors: the stirring of
the normal fluid is weaker, given that the vortex velocity is smaller ($\dot{\bolds} \sim 1/a$), and
the hydrodynamic interactions are less intense as vortices are further apart.
This last feature is characteristic of active fluid systems, as observed for drafting particles
in optical vortices \cite{Reichert2004}, for cyclists facing a steep hill where drafting is negligible \cite{Trenchard2016} and in 
the role played by ejection delay in the dispersion of fungal spores \cite{Roper2010}.
}
\begin{figure}[ht]
       \centering
       \includegraphics[width=.90\linewidth]{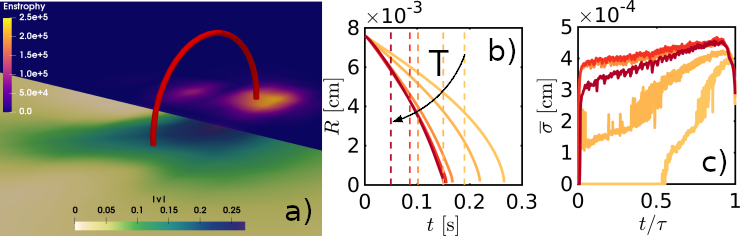}
       \caption{
       Coupled NS-VF model. Normal fluid flow disturbances at $T=1.95~\rm K$
       generated by a single superfluid vortex ring of initial
       radius $R_0=7.6 \times 10^{-3}\text{cm}$ traveling from left to
       right.
       \textbf{a)}: normal fluid's enstrophy $|\bm{\omega}_n|^2$ (in $\rm s^{-2}$, top)
       and speed $|\boldv_n|$ (in cm/s, bottom)
       plotted on the horizontal plane $z=0$ at time $t=0.16~\rm s$; the superfluid vortex ring is the red tube.
       \textbf{b)}~Temporal evolution of the radius $R$ (in cm) of an isolated vortex ring
       of initial radius $R_0=7.6 \times 10^{-3}\text{cm}$, moving in an initially quiescent normal fluid at
       temperatures $T=1.7,\,1.8,\,1.95,\,2.0,\,2.1~\rm K$ (from yellow to red)
       \textbf{c)}~Temporal evolution of the size $\bar{\sigma}$ of the normal fluid flow disturbances generated by the
       shrinking vortex ring; colors as in b);
       time $t$ is normalized by the vortex ring lifetime $\tau$.}
       \label{fig:side}
\end{figure} 

\blue{
To assess the role played by the coupling on its own, we study, employing 
the coupled NS-VF model, the 
dynamics of a single, isolated vortex ring of initial radius $R_0=7.6 \times 10^{-3}\,\text{cm}$, 
in an initially quiescent normal fluid.
Panel a) of  Fig.~\ref{fig:side} shows the isolated superfluid 
vortex ring (in red) traveling from left to right at $t=0.16~\rm s$
when the temperature $T=1.95~\rm K$. 
The normal fluid enstrophy density $|\bm{\omega}_{\rm n}|^2$ 
and the magnitude of the normal fluid velocity, $|\boldv_{\rm n}|$, 
are displayed respectively in the upper and lower halves
of the horizontal plane $z=0$ (perpendicular to the plane
containing the superfluid vortex ring). 
We observe two enstrophy structures which can be thought as 
two vortex rings in the normal fluid \cite{kivotides-barenghi-samuels-2000}.
Similar normal fluid enstrophy structures are also visible near all 
vortex lines in a bundle, see Fig.~\ref{fig:BundleVis3D} (bottom row).}
 
As the isolated superfluid vortex ring moves in the normal fluid and
perturbs it, it loses energy: its radius $R$ therefore shrinks with time,
as shown in Fig.~\ref{fig:side}(b), with corresponding 
lifetimes $\tau$ decreasing for increasing temperatures \cite{barenghi-donnelly-vinen-1983}. 
Lifetimes of vortex rings predicted by the coupled NS-VF model are roughly
twice the lifetimes predicted by Schwarz's VF model (indicated by
vertical dashed lines in Fig.~\ref{fig:side}(b)): \blue{the inclusion
of the coupling in the model indeed reduces the dissipation with respect to Schwarz's VF model. However,
if we compute the dissipation reduction $\chi (t)$ during
the shrinking of the rings (Fig.~\ref{fig:dissip.red}(b))), we observe that the dissipation
actually increases with respect to its initial value ($\chi < 0$). Hence, the dissipation reduction ($\chi \rightarrow 1$)
observed in the dynamics of bundles, responsible for their significantly enhanced lifetime (larger than $\tau_0$,
see Fig.~\ref{fig:BundleLengthAndEnergy}(a)), consistent with experimental measurements, 
uniquely stems from collective hydrodynamic cooperation.
}

\blue{By employing an enstrophy-weighed average approach (see Appendix \ref{calc.sigma}), we calculate the 
typical size $\bar{\sigma}$ of the normal fluid enstrophy structures
generated by vortex rings and report its temporal behaviour in Fig.~\ref{fig:side}(c). 
Subsequently we compute the dimensionless ratio $\ell/\bar{\sigma}$,
$\ell=a\sqrt{\pi/N}$ being the initial average intervortex spacing, obtaining
$0.93 \lesssim \ell/\bar{\sigma} \lesssim 1.7$ (simulations in Fig.~\ref{fig:BundleLengthAndEnergy})
and $1.7 \lesssim \ell/\bar{\sigma} \lesssim 6.8$ (simulations in Fig.~\ref{fig:dissip.red}(a)): 
this range of values assumed by $\ell/\bar{\sigma}$ implies that the normal fluid perturbations
indeed play a role in the vortex bundle dynamics, confirming the hydrodynamic
cooperative nature of the dissipation reduction observed.
}
%
\blue{
Interestingly, the values of $\bar{\sigma}$ are comparable to the size of  
solid hydrogen tracking particles used in current experiments, reinforcing 
recent suggestions ascribing the observed statistics 
of particle velocities also to the indirect interaction between particle and vortices, \textit{i.e.} 
via the disturbances generated in the normal 
fluid by superfluid vortices \cite{mastracci-etal-2019,svancara-etal-2021}. 
}


\section{Conclusions}
Using our coupled NS-VF model which takes 
in full account \cite{Galantucci2020}
the reciprocal interaction of the superfluid and the
normal fluid, we have found 
that a compact bundle of superfluid vortex rings creates a
disturbance in the normal fluid that is sufficiently strong to reduce
the \blue{overall} velocity difference between the two fluids, hence reduce
the friction on the superfluid vortex rings. 
While isolated superfluid vortex rings
quickly lose energy, shrink and vanish, we observe that bundled 
vortex rings remain
coherent and travel a significant distance compared to their size,
as observed in the experiments \cite{Borner1981,Borner1983,Borner1985}.
We have also found that the bundle remains coherent
in the limit of zero temperature (no normal fluid), 
again in agreement with experiments and previous works~\cite{Wacks2014}.

\blue{We show that the observed dissipation reduction in bundles is a collective effect
stemming from the hydrodynamic cooperation of vortices.} This cooperation is
similar to what has been observed in systems of active 
particles such as swimming
bacteria \cite{Lopez2015,Martinez2020,Guo2018bacteria}, 
fungal spores \cite{Roper2010}, 
racing cyclists \cite{Blocken2018,Belden2019}
and particle pairs trapped in an optical vortex
\cite{Reichert2004,Grujic2007}, in which self-organized structures emerge
from energy-saving mechanisms \cite{Trenchard2016}.
The system that we have investigated, superfluid helium, is however richer: 
whereas in fact in the cited active matter systems the agents, 
besides modifying the common background fluid, may interact with each 
others directly only via short-range two-body collisions, in our case 
vortex lines also experience a significant
collective long-range Biot-Savart interaction  
which, for instance, induces them to collectively rotate around each other (leapfrogging). 
Superfluid helium can hence be considered as a peculiar kind
of active fluid,
distinguished by a 4-way coupled dynamics
which potentially determines characteristics of turbulence 
in both superfluid and normal fluid components. 

The effect
of coupling and drag reduction on the statistics of superfluid turbulence
and on other integral quantities (such as helicity \cite{Galantucci2021} if
for instance the initial bundle is twisted)
will be the topic of future research, as well as the implications
for vortex dynamics in the more viscous helium isotope $^3$He.

\begin{acknowledgments}
\emph{Acknowledgments.}
We are grateful to the Royal Society for supporting
this project (award n. IES \textbackslash R2\textbackslash 181176).
LG and CFB acknowledge the support of the Engineering and Physical Sciences Research Council 
(Grant No. EP/R005192/1).  LG acknowledges the support of Istituto Nazionale di Alta Matematica (INdAM).
GK was supported by the Agence Nationale de la Recherche (project GIANTE ANR-18-CE30-0020-01).
Computations were carried out at the Mesocentre SIGAMM, hosted at the Observatoire de la Cote d'Azur,
and at the HPC Rocket Cluster at Newcastle University.
\end{acknowledgments}

\appendix

\section{Calculation of $\bar{\sigma}(t)$}
\label{calc.sigma}
The centre $\mathbf{x}_\omega(t)$ of the normal fluid enstrophy distribution 
$\displaystyle |\bm{\omega}_{\rm n}(\mathbf{x},t)|^2$ on the $z=0$ plane is calculated as follows,
\begin{equation}
\mathbf{x}_\omega(t) =
\frac{\displaystyle\int \!  \! \int \mathbf{x} \; |\bm{\omega}_{\rm n}(\mathbf{x},t)|^2 \; d \mathbf{x}}{\displaystyle\int  \! \! \int |\bm{\omega}_{\rm n}(\mathbf{x},t)|^2 \; d \mathbf{x}} \;\; , \label{eq:vorticity_center}
\end{equation}
where $\mathbf{x}_\omega(t)=\left ( x_\omega(t) \; , \; y_\omega(t) \; , \; 0 \right )$ and 
$\mathbf{x}=\left ( x \; , \; y \; , \; 0\right )$ as the calculation is performed on the $z=0$ plane.\\
The normal fluid vortex size $\bar{\sigma}$ whose temporal evolution is reported in Fig.~\ref{fig:side} (b) is given by
$\bar{\sigma}(t)=\sqrt{R_n(t) \; R_t(t)}$ where $R_n$ and $R_t$ are computed as follows,
\begin{eqnarray}
R_n^2(t) & = &
\frac{\displaystyle\int \!  \! \int \left [ (\mathbf{x}-\mathbf{x}_\omega(t)\,)\cdot \hat{\mathbf{n}}(t)\,\right ]^2 \; |\bm{\omega}_{\rm n}(\mathbf{x},t)|^2 \; d \mathbf{x}}{\displaystyle\int  \! \! \int |\bm{\omega}_{\rm n}(\mathbf{x},t)|^2 \; d \mathbf{x}} \;\; , \label{eq:R_n} \\[2mm]
R_t^2(t) & = &
\frac{\displaystyle\int \!  \! \int \left [ (\mathbf{x}-\mathbf{x}_\omega(t)\,)\cdot \hat{\mathbf{t}}(t)\,\right ]^2 \; |\bm{\omega}_{\rm n}(\mathbf{x},t)|^2 \; d \mathbf{x}}{\displaystyle\int  \! \! \int |\bm{\omega}_{\rm n}(\mathbf{x},t)|^2 \; d \mathbf{x}} \;\; , \label{eq:R_t}
\end{eqnarray}
where $\hat{\mathbf{t}}(t)$ is the unit vector indicating the direction of the mutual friction force $\mathbf{f}_{\rm ns} (t)$
exerted by the vortex ring onto the normal fluid on the $z=0$ plane,
and $\hat{\mathbf{n}}(t)$ is the orthogonal direction to  $\hat{\mathbf{t}}(t)$ lying on $z=0$ plane.

\section{The NS-VF model\label{App:method}}

The numerical methods used to implement Schwarz's VF and the coupled NS-VF models 
are described in detail in Ref.~\cite{Galantucci2020}. Here we summarize the main 
characteristics of the coupled NS-VF model.

\subsection{Superfluid vortex tangle and normal fluid velocity field evolution}
The temporal evolution of the superfluid vortex tangle $\mathcal{L}$ is performed 
employing the well-established Lagrangian VF method elaborated by Schwarz \cite{Schwarz1988,hanninen-baggaley-2014}
which discretizes vortex lines in a finite number of line elements whose equation of motion is
given by Eq.~(1) in the main manuscript. The singularity of the Biot-Savart integral, Eq.~(2) in the main manuscript, is regularized 
by taking into account the finite size of the vortex core \cite{Schwarz1988}. We compute the full Biot-Savart integral (no tree-approximation).
As reconnections are not intrinsically predicted by the VF method,
an additional algorithm has to be employed changing the topology of the vortex configuration when
two vortex lines become closer than a set threshold. 

The evolution of the normal fluid velocity field $\mathbf{v}_{\rm n}$
is computed integrating the Navier-Stokes equations (Eqs.~(3)~-~(4) in main manuscript) using
a standard pseudo-spectral code de-aliased employing the 2/3-rule.
We refer to established literature for further details concerning the standard
algorithm employed for the numerical integration of the Navier Stokes equations \cite{gottlieb1977numerical}.

\subsection{Mutual friction force}
The distinguishing features of our
coupled NS-VF algorithm actually concern the
modeling of the mutual friction force per unit length $\mathbf{f}_{\rm ns}$ 
in Eq.~(4) of the main manuscript. We describe the interaction between superfluid
vortices and the normal fluid employing a classical low Reynolds number
approach \cite{proudman-pearson-1957}
revisiting a recent
framework used in superfluid turbulence \cite{kivotides-2018}. According
to this approach, the mutual friction force which the superfluid vortices
exert on the normal fluid reads as follows,

\begin{equation}
\mathbf{f}_{\rm ns} [\mathbf{s}] = - D \,\mathbf{s}' \times \left [ \mathbf{s}' \times 
\left (\dot{\mathbf{s}} - \mathbf{v}_{\rm n}  \right ) \right ] - 
\rho_n\kappa\; \mathbf{s}' \times \left (\dot{\mathbf{s}} - \mathbf{v}_{\rm n} \right ) \;\; ,
\label{eq:fns}
\end{equation}
where the drag coefficient $D=D[\mathbf{s}]$ is 
\begin{equation}
\displaystyle D= \frac{4\pi\rho_{\rm n}\nu_{\rm n}}{[\frac{1}{2}-\gamma-\ln\left 
( \frac{|\mathbf{v}_{\rm n_{\perp}} - \dot{\mathbf{s}}|a_0}{4\nu_{\rm n}}\right )]},
\label{eq:Drag_D}
\end{equation}
$\gamma=0.5772$ being the Euler-Mascheroni constant,
$\mathbf{v}_{\rm n}$ is evaluated on the vortex lines, that is to say
$\mathbf{v}_{\rm n}=\mathbf{v}_{\rm n}[\mathbf{s}]$ (the interpolation being performed using
fourth-order \emph{B-splines}),
and the quantity
$\mathbf{v}_{\rm n_{\perp}}$
indicates the component of the normal fluid velocity 
lying on a plane orthogonal to $\mathbf{s}'$.
The use of the expression (\ref{eq:fns}) for $\mathbf{f}_{\rm ns}$ leads to a recalculation 
of friction coefficients $\alpha$ and $\alpha'$ in Eq.~(1) of the main manuscript, as reported in the next section.

As the mutual friction force $\mathbf{f}_{\rm ns}$ is $\delta$-supported on the vortex lines,
its numerical distribution on the Eulerian computational grid where we compute the normal fluid velocity $\mathbf{v}_{\rm n}$
must be handled with care, in order to avoid spurious numerical artifacts.  
To address this issue, we adopt the same
rigorous regularization approach which has been used 
to take into account
the strongly localized response of active point-like 
particles 
in classical turbulence
\cite{gualtieri_picano_sardina_casciola_2015}. 
The advantage of adopting this method is that the regularization of the 
exchange of momentum between point-like active agents and viscous flows 
is based on the physics of the generation of vorticity 
and its viscous diffusion at very small scales. 
In our case, the justification for the use of this model arises from
the very small Reynolds numbers characterizing the normal fluid 
disturbances generated by the moving superfluid vortices ($ \rm Re \approx 10^{-5} \div 10^{-4}$).

\subsection{Calculation of friction coefficients in the coupled NS-VF model \label{App:mutualFrictionCoeff}}

Here we briefly describe the derivation of the expression of the mutual friction coefficients
in the coupled NS-VF model (for further details, the reader is referred to Ref.~\cite{Galantucci2020}). 
The starting point is the classical, low Reynolds number theoretical approach which we employ
to model the mutual friction force. This framework leads to Eq.~(4) in the main manuscript accounting for the force per 
unit length $-\mathbf{f}_{\rm ns}$ which the normal fluid exerts onto the 
superfluid vortices. The \red{superfluid vortices} also suffer
a Magnus force $\mathbf{f}_{\rm M}$ as they are immersed in an 
inviscid fluid (the \red{superfluid}) surrounded by a circulation
and in relative motion with respect to the superfluid itself. The expression of the Magnus force per unit length 
exerted onto the superfluid vortices is as follows, 
\begin{equation}
\displaystyle
\mathbf{f}_{\rm M} = \rho_{\rm s} \kappa \; \mathbf{s}' \times \left ( \dot{\mathbf{s}} - \mathbf{v}_{\rm s} \right ) \; \; .
\label{eq:f_M}
\end{equation}
Since the vortex core is much smaller then any
other scales of the flow, the vortex inertia can be neglected and
as a consequence the sum of all
forces acting on the vortices vanishes, {\it i.e.} 
$\mathbf{f}_{\rm M} -\mathbf{f}_{\rm ns} = 0$.
Assuming that each vortex line element moves orthogonally to its unit tangent vector, 
\textit{i.e.} $\dot{\mathbf{s}} \cdot \mathbf{s}' = 0$, the balance of forces leads to the
following equation of motion,
\begin{equation}
\displaystyle
\dot{\mathbf{s}} = \mathbf{v}_{\rm s_{\perp}}
+ \beta \mathbf{s}' \times  \left ( \mathbf{v}_{\rm n} - \mathbf{v}_{\rm s} \right )
+ \beta' \mathbf{s}' \times \left [ \mathbf{s}' \times \left ( \mathbf{v}_{\rm n} - \mathbf{v}_{\rm s} \right ) \right ] \; \; ,
\label{eq:vortex_motion_final}
\end{equation}
where $\mathbf{v}_{s_{\perp}}$ 
indicates the component of the superfluid velocity 
lying on a plane orthogonal to $\mathbf{s}'$
and $\beta$ and $\beta'$ are the redetermined mutual friction coefficients for the coupled model.

The expressions for $\beta$ and $\beta'$ are as follows,
\begin{eqnarray*}
  \beta =  \frac{a}{(1+b)^2 + a^2} > 0,&\quad&\beta' = - \frac{b(1+b) + a^2}{(1+b)^2 + a^2} < 0
\end{eqnarray*} 

%
where 
\begin{equation*}
a=\frac{D}{\rho_{\rm s} \kappa}= 4\pi\;
\left(\frac{\rho_{\rm n}} {\rho_{\rm s}}\right) 
\left( \frac{\nu_{\rm n}}{\kappa} \right) \;
\frac{1}{[\frac{1}{2}-\gamma-\ln\left ( \frac{|\mathbf{v}_{\rm n_{\perp}} - \dot{\mathbf{s}}|a_0}{4\nu_{\rm n}}\right )]}\;\;
\end{equation*}
and $b=\displaystyle\frac{\rho_{\rm n}}{\rho_{\rm s}}$.
Thus, from the physical point of view, the motion of the vortices 
is governed only by temperature amd pressure,
determining $\rho_{\rm n}/\rho_{\rm s}$ and $\nu_{\rm n}/\kappa$, and
the normal fluid Reynolds number $\rm Re = |\mathbf{v}_{\rm n_{\perp}} - \dot{\mathbf{s}}| a_0/\nu_{\rm n} $. 
In the numerical simulations, we employ values of the densities $\rho_{\rm n}$ and $\rho_{\rm s}$ and of the 
normal fluid kinematic viscosity $\nu_{\rm n}$ consistent with temperature $T=1.95$K at saturated vapor pressure
\cite{donnelly-barenghi-1998}. Correspondingly, also the values of the friction coefficients $\alpha$ and $\alpha'$
employed in the Schwarz VF model are consistent with experimental values \cite{donnelly-barenghi-1998}.

\section{Physical and numerical parameters}\label{App:Params}
\subsection{Superfluid vortex tangle simulations \label{App:ParamBundle}}

Following the VF model elaborated by Schwarz \cite{Schwarz1988,hanninen-baggaley-2014}, we discretize the vortex tangle
$\mathcal{L}$ in a set of $N_p$ vortex line elements centered in $\mathbf{s}_{i}(t)=\mathbf{s}(\xi_i,t)\; , \; i=1, \dots N_p$ ,
where $\xi_i = i \Delta \xi$ is the discretized arclength with discretization $\Delta \xi \in [\delta \; , \; 2\delta]$ where
$\delta = 4.0 \times 10^{-4} \rm cm$. 
The normal fluid is solved on a three-dimensional computational grid with 
$ \{ N_x \; , \; N_y \; , \; N_z\} = \{ 512 \; , \; 512 \; , \; 512 \}$ collocation points in each cartesian direction. The 
computational domain is a periodic box of size 
$L_x \times L_y \times L_z$ with $L_x=L_y=L_z=10^{-1}\rm cm$ which leads to grid spacings  
$\Delta x = \Delta y = \Delta z = 1.95 \times 10^{-4}\rm cm$. The size of the computational box is identical for the 
calculation of the vortex filaments  ($L_x=L_y=L_z=10^{-1}\rm cm$) and also in this calculation 
we use periodic boundary conditions. 
The time step $\Delta t$ employed in the computation of the temporal evolution of the `large' bundle 
($N=169$, cf. main manuscript) is $\Delta t = 5.0 \times 10^{-6} \rm s$, while the time step used for the `thin' bundle ($N=37$)
is $\Delta t = 6.25 \times 10^{-6} \rm s$. In order to distribute the mutual friction force 
$\mathbf{F}_{\rm ns}$ over the computational grid where the normal fluid velocity is resolved, 
before employing the regularization adopted in classical turbulence \cite{gualtieri_picano_sardina_casciola_2015}, 
we interpolate the vortex filaments with a cubic kernel over an arc-length sub-scale $\Delta \xi/4$. We validated this interpolation
method on the motion of individual vortex rings. 

\subsection{Single superfluid vortex ring simulations \label{App:ParamRings}}

In this set of simulations whose results are summarized in Fig.~\ref{fig:dissip.red}(b)  and \ref{fig:side}, 
we use a finer discretization
of the vortex lines, $\delta$ being equal to $8.0 \times 10^{-5} \rm cm$. This results in a smaller time step 
$\Delta t_{\rm VF} = 6.25 \times 10^{-7} \rm s$. For the normal fluid velocity computation we use 
$ \{ N_x \; , \; N_y \; , \; N_z\} = \{ 256 \; , \; 256 \; , \; 256 \}$ collocation points in each cartesian direction leading to 
$\Delta x = \Delta y = \Delta z = 3.90 \times 10^{-4}\rm cm$, as the computational periodic 
box is $L_x \times L_y \times L_z$ with $L_x=L_y=L_z=10^{-1}\rm cm$. The time step employed for the calculation of the 
normal fluid velocity field is $\Delta t_{\rm NS} = 2.50 \times 10^{-5} \rm s$.

\end{document}